# On the Input-Output Monotonicity of Voltage Dynamics of Power System with Grid-Forming Converters

Zhenyao Li, Shengwen Liao, Qian Zhang, Xuechun Zhang, Deqiang Gan

*Abstract*—**Integration of renewable resources is profoundly reshaping the dynamics of modern power systems. This study shows that the voltage dynamics of power systems with multiple grid-forming (GFM) converters often enjoys a desirable property called input-output monotonicity. A systematic approach for computing the derivatives of the voltage subsystem is presented first, which provides insight into the structural characteristics of these models. Next, the sign pattern of the trajectory Jacobian matrix associated with the voltage subsystem is analyzed and revealed. The analysis indicates that the voltage dynamics of power systems often exhibits the so-called input-output monotonicity property. The theoretical results are then validated through several simulation examples, underscoring their practical implications.**

## I. Introduction

With the high penetration of inverter-based resources into modern power system, the dynamic behavior of power systems is becoming more complex and to some extent more vulnerable [1]-[3]. One of the issues that has caused wide-spread concern in power community is the weakened voltage control capability in such modern power systems.

Traditionally, there exists a portfolio of voltage control strategies in power systems. Synchronous machines and SVCs offer continuous responses to reactive power variations, while under load tap changer and capacitor banks are equipped to provide discrete responses. As the last resort, under voltage load shedding emergency control can be activated to prevent voltage instability or collapse.

To meet the voltage control challenge in power systems with high penetration of inverter-based resources, a few innovative control strategies have been proposed in the literature. A combined active and reactive power control strategy based on model predictive control was proposed in [4] as a means for improving wind farm voltage control. In [5], a decentralized real-time adaptive under voltage load shedding scheme which combines fuzzy logic control and particle swarm optimization was developed. This scheme was designed to address the issue of fault-induced delayed voltage recovery and short-term voltage instability. The dynamic voltage support capability of photovoltaic systems was exploited and voltage stability was improved using active and reactive power injection in [6]. Singular value decomposition and reconstructed power flow matrix were used in [7] to track the critical point of voltage collapse in renewable energy generations integrated power systems. A model predictive control was proposed in [8] to enhance the dynamic performance of system voltage.

Grid-forming (GFM) converters have also been developed to cope with the afore-mentioned voltage control issue [9]-[11]. GFM converters are equipped with virtual inertia, voltage regulation function and damping effect to simulate synchronous generators (SGs) [12]-[14]. Therefore, the dynamic behavior of a GFM converter is similar to that of a controllable voltage source [15]. Since a GFM converter can rapidly change its output power in response to network conditions, it can therefore actively participate in voltage control of power systems [16]. While the basic conception of GFM control is well-understood, much less is known about how multiple GFM converters and SGs interact in modern power systems. This is the main thrust of this work.

Our recent research [17] shows that the trajectory Jacobian matrix (i.e. the Jacobian matrix along the trajectory) of a power system exhibits apparent sign pattern, this permitted us to study voltage dynamics of power systems using a divide-and-conquer strategy. The goal of this work is to extend the investigation to systems with GFM converters, with a distinct focus on the property of voltage dynamics. Specifically, the paper reveals that the voltage trajectories of a power system under normal operation possess a desirable property called input-output monotonicity.

The remaining of the paper is organized as follows. First, the sign pattern of the voltage subsystem with GFM converters is deduced using a systematic approach. Then, the input-output monotonicity of voltage dynamics is analyzed based on the monotone system theory. Finally, the validity of the above results is verified with several examples, and the conclusions are draw.

## II. The Model of Power System and A Systematic Approach to Compute The Trajectory Derivatives

In this section, the model of the power system under study is given, and a systematic approach for computing the

[1]This work is supported by National Natural Science Foundation of China (Transient stability analysis and control of AC/DC power grids with high penetration of renewable power generation, U2166601).

The authors are with the College of Electrical Engineering, Zhejiang University, Hangzhou, China. (Emails: {12110085, liaoshengwenee, zhangqianleo, 12310048, dgan}@zju.edu.cn)

trajectory derivatives is developed.

### A. The Model of SGs

The single-axis model for SGs is as follows [18]

$$\boldsymbol{f_1}: \begin{cases} \dot{\delta} = \omega_0(\omega - 1) \\ M\dot{\omega} = P_m - P_e - D(\omega - 1) \\ T'_{d0}\dot{E}'_q = E_{fd} - E'_q - (x_d - x'_d)I_d \\ T_A\dot{E}_{fd} = K_A(V_{ref} - |V_t|) - E_{fd} \end{cases} \quad (1)$$

where $\delta$ is the rotor angle, $\omega$ and $\omega_0$ are the frequency of machines and the synchronous frequency, $M$ is the inertia constant, $P_m$ and $P_e$ are mechanical power and electromagnetic power, $D$ is the damping coefficient, $T'_{d0}$ is the open-circuit time constant, $E'_q$ is the $q$-axis component of the voltage behind transient reactance, $E_{fd}$ is the excitation voltage, $x_d$ and $x'_d$ are the $d$-axis synchronous reactance and the $d$-axis transient reactance, $I_d$ is the $d$-axis current, $T_A$ is the time constant, $K_A$ is the excitation gain, $V_{ref}$ is the reference voltage, $|V_t|$ is the amplitude of terminal voltage.

As is well-known, the dynamic behavior of power systems is extremely complex. A viable method is to decompose the system into several subsystems and analyze the structural properties of subsystems [19]. When studying the dynamics of voltage, one is often only concerned with the subsystems such as the voltage behind transient reactance and the excitation system of SGs, while the motion of the rotor angles can be effectively ignored without imposing significant errors [20]. This leads us to consider the dynamic equations of the transient voltage subsystem of the SGs:

$$\begin{cases} T'_{d0}\dot{E}'_q = E_{fd} - E'_q - (x_d - x'_d)I_d \triangleq F_{Eq} \\ T_A\dot{E}_{fd} = K_A(V_{ref} - |V_t|) - E_{fd} \triangleq F_{Efd} \end{cases} \quad (2)$$

Using the stator voltage equation:

$$I_d = x_d'^{-1}(E'_q - V_q) \quad (3)$$

where $V_q$ is the $q$-axis component of the terminal voltage, and the coordinate transformation equation:

$$V_q = 0.5(V_t + \bar{V}_t)\cos\delta - 0.5j(V_t - \bar{V}_t)\sin\delta \quad (4)$$

where $\delta$ is the rotor angle, we can get the dynamic equations with only terminal voltage and the state variables:

$$\begin{cases} T'_{d0}\dot{E}'_q = (x_d x_d'^{-1} - \boldsymbol{I})V_q + E_{fd} - x_d x_d'^{-1}E'_q = F_{Eq} \\ T_A\dot{E}_{fd} = K_A(V_{ref} - |V_t|) - E_{fd} = F_{Efd} \end{cases} \quad (5)$$

where $\boldsymbol{I}$ is the identity matrix.

The output power of $i$-th SG can be obtained as follows [21] :

$$\begin{aligned} S_{SG,i} = g_{SG}(E'_{q,i}, V_{ti}) &= 0.5je^{-j2\delta_i}V_{ti}^2(x_{q,i}^{-1} - x_{d,i}'^{-1}) \\ &+ jV_{ti}x_{d,i}'^{-1}E'_{q,i}e^{-j\delta_i} - 0.5j(x_q^{-1} + x_d'^{-1})V_{ti}\bar{V}_{ti} \end{aligned} \quad (6)$$

where $S_{SG}$ is the output power of SGs, $x_q$ is the $q$-axis synchronous reactance.

### B. The Model of GFM Converters

Virtual synchronous machine (VSM) represents a mainstream control strategy for GFM converters at present. This paper presents a case study of VSM as a representative example, and the same methodology may be employed to validate other control strategies.

The model of GFM converters under VSM strategy is illustrated in Fig. 1 below [22].

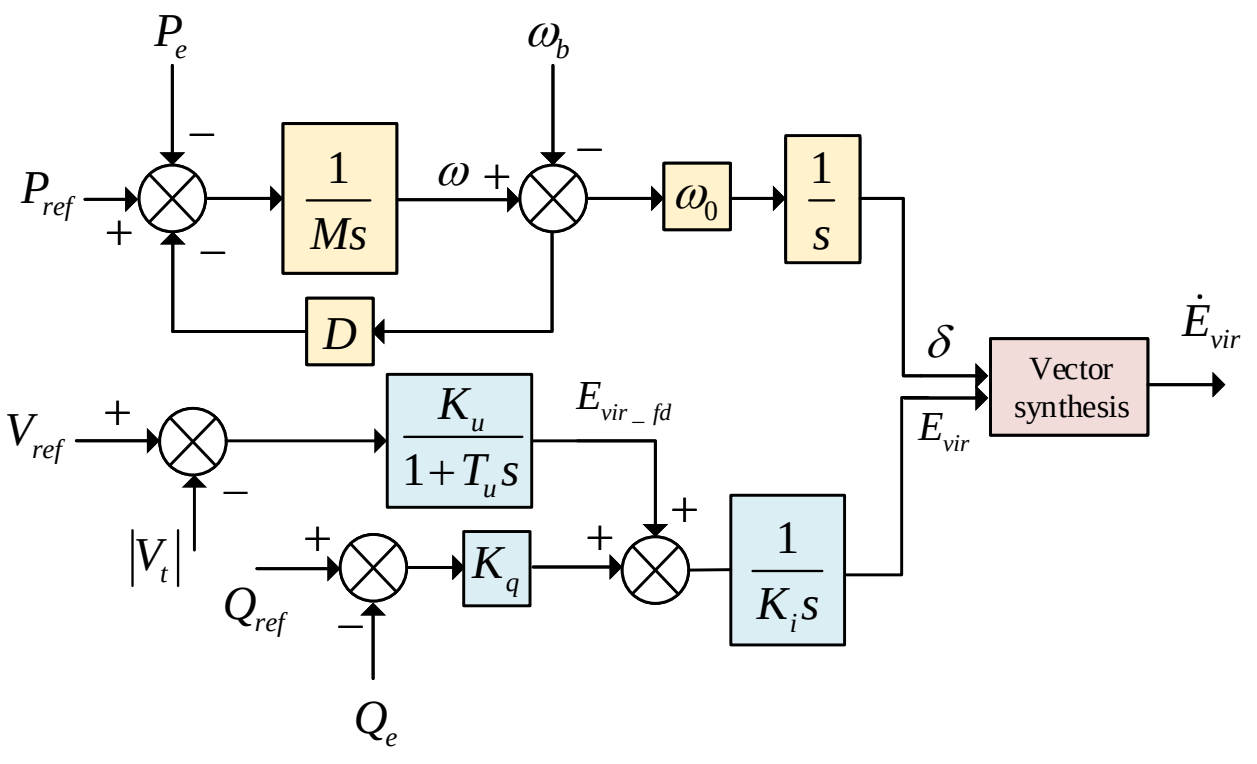

Fig. 1. Control block diagram of VSM strategy

After ignoring the motion of the angles, the mathematical model of GFM converters according to the control block diagram depicted in Fig. 1 can be obtained as:

$$\begin{cases} K_i\dot{E}_{vir} = E_{vir_fd} + K_q(Q_{ref} - Q_e) \triangleq F_{vir} \\ T_u\dot{E}_{vir_fd} = K_u(V_{ref} - |V_t|) - E_{vir_fd} \triangleq F_{vir_fd} \end{cases} \quad (7)$$

where $E_{vir}$ is the internal voltage of the converter, $K_i$ and $T_u$ are the time constants, $E_{vir_fd}$ is the virtual excitation voltage, $Q_{ref}$ and $Q_e$ are the reference reactive power and output reactive power of a GFM converter, $K_q$ and $K_u$ are the power coefficient and voltage coefficient in the $Q$-$V$ droop control, respectively.

The output power $S_{GFM}$ of the $i$-th GFM converter is:

$$S_{GFM,i} = g_{GFM}(E_{vir,i}, V_{ti}) = je^{-j\delta_i}V_tE_{vir,i}x_{l,i}^{-1} - jx_{l,i}^{-1}V_{ti}\bar{V}_{ti} \quad (8)$$

where $x_l$ is the outlet impedance of the converter.

### C. A Systematic Approach to Compute Derivatives

In our recent work [17], we studied the sign pattern of a power system model, assuming that loads are represented by constant impedances. In this section this restriction is removed by adopting a systematic approach for computing derivatives.

Notice that the model of power systems typically includes three parts as follows.

1) Dynamic equation of power system components:

$$\dot{\boldsymbol{x}} = \boldsymbol{F}(\boldsymbol{x}, \boldsymbol{V}) \quad (9)$$

where $\boldsymbol{x}$ is the state variables of the components, $\boldsymbol{V}$ is the complex terminal voltage vector of components, $\boldsymbol{F}$ is the dynamic model of the components. Notice that both model (5) and (7) can be written in this form.

2) Network equation:

$$\boldsymbol{YV} = \bar{\boldsymbol{S}}./\bar{\boldsymbol{V}} \quad (10)$$

where $\boldsymbol{S}$ is the complex power injected into the power system, $\boldsymbol{Y}$ is the admittance matrix, the over-line denotes the conjugation operation, and "./" denotes pointwise division, following the standard MATLAB convention.

3) Interface equation of power injection components:

$$\boldsymbol{S} = \boldsymbol{g}(\boldsymbol{x}, \boldsymbol{V}) \quad (11)$$

The output powers of SGs and GFM converters, refer to equations (6) and (8), can be written in this form.

The Jacobian matrix of the power system can be obtained by employing the chain rule of derivatives:

$$\frac{\partial \dot{\boldsymbol{x}}}{\partial \boldsymbol{x}}=\frac{\partial \boldsymbol{F}}{\partial \boldsymbol{V}}\frac{\partial \boldsymbol{V}}{\partial \boldsymbol{x}}+\frac{\partial \boldsymbol{F}}{\partial \bar{\boldsymbol{V}}}\frac{\partial \bar{\boldsymbol{V}}}{\partial \boldsymbol{x}}+\frac{\partial \boldsymbol{F}}{\partial \boldsymbol{x}} \tag{12}$$

where $\partial \boldsymbol{F}/\partial \boldsymbol{V}$ and $\partial \boldsymbol{F}/\partial \boldsymbol{x}$ can be obtained from the dynamic equation of power components. Matrix $\partial \boldsymbol{V}/\partial \boldsymbol{x}$ can be obtained by combining the network equation and interface equation as:

$$\boldsymbol{YV}=\bar{\boldsymbol{g}}(\boldsymbol{x},\boldsymbol{V})./\bar{\boldsymbol{V}} \tag{13}$$

Differentiating both sides of the above equation with respect to $\boldsymbol{x}$, one can get:

$$\mathrm{diag}(\bar{\boldsymbol{V}})\boldsymbol{Y}\frac{\partial \boldsymbol{V}}{\partial \boldsymbol{x}}=\frac{\partial \bar{\boldsymbol{g}}}{\partial \boldsymbol{V}}\frac{\partial \boldsymbol{V}}{\partial \boldsymbol{x}}+\frac{\partial \bar{\boldsymbol{g}}}{\partial \bar{\boldsymbol{V}}}\frac{\partial \bar{\boldsymbol{V}}}{\partial \boldsymbol{x}}+\frac{\partial \bar{\boldsymbol{g}}}{\partial \boldsymbol{x}}-\mathrm{diag}(\bar{\boldsymbol{S}}./\bar{\boldsymbol{V}})\frac{\partial \bar{\boldsymbol{V}}}{\partial \boldsymbol{x}} \tag{14}$$

where $\mathrm{diag}(\cdot)$ is the function of a diagonal matrix consisting of elements in parentheses. Introduce the following variable substitutions:

$$\begin{cases}\boldsymbol{A}=\mathrm{diag}(\bar{\boldsymbol{V}})\boldsymbol{Y}-\dfrac{\partial \bar{\boldsymbol{g}}}{\partial \boldsymbol{V}}\\ \boldsymbol{B}=\mathrm{diag}(\bar{\boldsymbol{S}}./\bar{\boldsymbol{V}})-\dfrac{\partial \bar{\boldsymbol{g}}}{\partial \bar{\boldsymbol{V}}}\\ \boldsymbol{C}=\dfrac{\partial \bar{\boldsymbol{g}}}{\partial \boldsymbol{x}}\end{cases} \tag{15}$$

Substituting (15) into (14) to get:

$$\begin{cases}\boldsymbol{A}\dfrac{\partial \boldsymbol{V}}{\partial \boldsymbol{x}}+\boldsymbol{B}\dfrac{\partial \bar{\boldsymbol{V}}}{\partial \boldsymbol{x}}=\boldsymbol{C}\\ \bar{\boldsymbol{A}}\dfrac{\partial \bar{\boldsymbol{V}}}{\partial \boldsymbol{x}}+\bar{\boldsymbol{B}}\dfrac{\partial \boldsymbol{V}}{\partial \boldsymbol{x}}=\bar{\boldsymbol{C}}\end{cases} \tag{16}$$

Solving (16) one can get:

$$\frac{\partial \boldsymbol{V}}{\partial \boldsymbol{x}}=[\boldsymbol{A}-\boldsymbol{B}\bar{\boldsymbol{A}}^{-1}\bar{\boldsymbol{B}}]^{-1}[\boldsymbol{C}-\boldsymbol{B}\bar{\boldsymbol{A}}^{-1}\bar{\boldsymbol{C}}] \tag{17}$$

Finally, the Jacobian matrix of the power system can be obtained by substituting (17) into (12).

To sum up, the computational steps of the above systematic approach is shown in Fig. 2. It is thought that the suggested procedure is flexible enough to accommodate a much wider class of power system models, in addition it has the advantage of being pipelined.

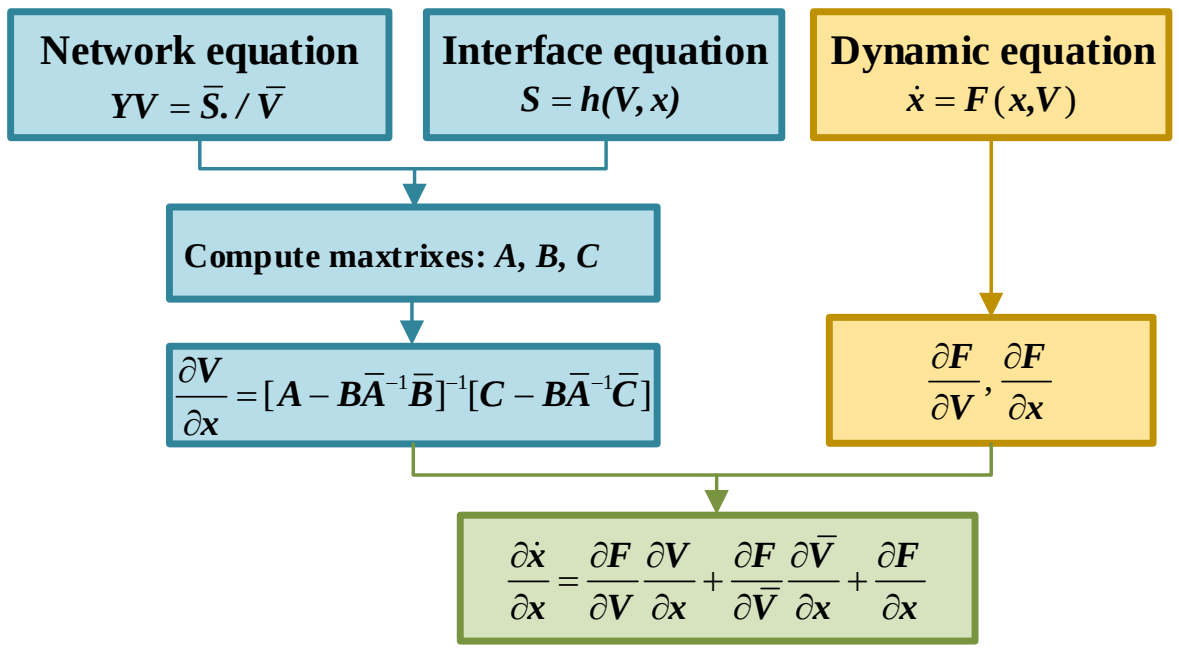

Fig. 2. Computational steps of the systematic approach for derivatives

## III. Sign Pattern of Transient Voltage Subsystem

This section gives a study on the sign pattern of Jacobian matrix for systems with SGs and GFM converters based on the systematic approach mentioned above.

### *A. Sign Pattern of SGs Voltage Dynamics*

In what follows we investigate the sign pattern of Jacobian matrices along trajectories, which play a decisive role in determining the dynamic voltage behavior of power systems.

#### *1) The Partial Derivative Matrix* $\partial \dot{\boldsymbol{E}}'_q/\partial \boldsymbol{E}'_q$

According to Fig. 2, the partial derivative of the voltage behind transient reactance dynamic equations of the SGs are obtained first:

$$\begin{cases}\dfrac{\partial \boldsymbol{F}_{Eq}}{\partial \boldsymbol{V}}=\mathrm{conj}(\dfrac{\partial \boldsymbol{F}_{Eq}}{\partial \bar{\boldsymbol{V}}})=\mathrm{diag}\left(\dfrac{x_{d,i}x'^{-1}_{d,i}-1}{2}e^{-j\delta_i}\right)\\ \dfrac{\partial \boldsymbol{F}_{Eq}}{\partial \boldsymbol{E}'_q}=\mathrm{diag}\left(-\dfrac{x_{d,i}}{x'_{d,i}}\right)\end{cases} \tag{18}$$

According to the interface equation (6), it follows that:

$$\begin{cases}\dfrac{\partial g_{SG,i}}{\partial V_i}=0.5j(x_{q,i}^{-1}+x'^{-1}_{d,i})\bar{V}_{it}\\ \dfrac{\partial g_{SG,i}}{\partial \bar{V}_i}=0.5j(x_{q,i}^{-1}+x'^{-1}_{d,i})V_{ti}-jx'^{-1}_{d,i}E'_{q,i}e^{j\delta_i}-je^{j2\delta_i}\bar{V}_{ti}(x_{q,i}^{-1}-x'^{-1}_{d,i})\\ \dfrac{\partial g_{SG,i}}{\partial E'_{q,i}}=-jx'^{-1}_{d,i}\bar{V}_{ti}e^{j\delta_i}\end{cases} \tag{19}$$

Substituting (19) into (15), one can get matrixes $\boldsymbol{A}$, $\boldsymbol{B}$ and $\boldsymbol{C}$ as follows:

$$\begin{cases}\boldsymbol{A}=\mathrm{diag}(\bar{\boldsymbol{V}})\boldsymbol{Y}-\mathrm{diag}(0.5j(x_{q,i}^{-1}+x'^{-1}_{d,i})\bar{V}_i)=\mathrm{diag}(\bar{\boldsymbol{V}})\boldsymbol{Y}_{\mathrm{aug}}\\ \boldsymbol{B}=\mathrm{diag}([0.5je^{j2\delta_i}\bar{V}_i(x_{q,i}^{-1}-x'^{-1}_{d,i})])\\ \boldsymbol{C}=\mathrm{diag}([-j\bar{V}_i x'^{-1}_{d,i}e^{j\delta_i}])\end{cases} \tag{20}$$

where

$$\boldsymbol{Y}_{\mathrm{aug}}=\boldsymbol{Y}-\mathrm{diag}(0.5j(x_{q,i}^{-1}+x'^{-1}_{d,i})) \tag{21}$$

According to the properties of the network admittance matrix, the elements of $\boldsymbol{A}$ are on the order of hundreds. So the elements of the conjugate inverse matrix are very small, and the order of magnitude is between $10^{-3}$ and $10^{-2}$. Typically, the order of magnitude of $\boldsymbol{B}$ is between 1 and 10. Therefore, the following approximate formula can be obtained:

$$\begin{cases}[\boldsymbol{A}-\boldsymbol{B}\bar{\boldsymbol{A}}^{-1}\bar{\boldsymbol{B}}]^{-1}\approx\boldsymbol{A}^{-1}=\boldsymbol{Y}_{\mathrm{aug}}^{-1}\mathrm{diag}(\bar{\boldsymbol{V}}^{-1})\\ \boldsymbol{C}-\boldsymbol{B}\bar{\boldsymbol{A}}^{-1}\bar{\boldsymbol{C}}\approx\boldsymbol{C}\end{cases} \tag{22}$$

Substituting (22) into (17), a simplified formula is obtained:

$$\begin{aligned}\frac{\partial \boldsymbol{V}}{\partial \boldsymbol{E}'_q}&=[\boldsymbol{A}-\boldsymbol{B}\bar{\boldsymbol{A}}^{-1}\bar{\boldsymbol{B}}]^{-1}[\boldsymbol{C}-\boldsymbol{B}\bar{\boldsymbol{A}}^{-1}\bar{\boldsymbol{C}}]\approx\boldsymbol{A}^{-1}\boldsymbol{C}\\ &=\boldsymbol{Y}_{\mathrm{aug}}^{-1}\mathrm{diag}([-jx'^{-1}_{d,i}e^{j\delta_i}])\\ &=\boldsymbol{Z}\left(\mathrm{diag}([x'^{-1}_{d,i}\cos\delta_i])+j\mathrm{diag}([x'^{-1}_{d,i}\sin\delta_i])\right)\end{aligned} \tag{23}$$

By adjusting the phase angle of the slack bus, the angle of the $q$-axis of the SGs relative to the reference angle can be located between ±45°, so that the real part of $\partial \boldsymbol{V}/\partial \boldsymbol{E}'_q$ is larger than the imaginary part. Since the admittance matrix is approximately Laplacian [23], it can be known that $\boldsymbol{Z}$ is approximately non-negative, neglecting the operator "$j$".

Meanwhile, $\boldsymbol{Z}$ is a diagonally dominant matrix which is a useful property that will be exploited in later sections. Therefore, the derivative of the dynamic equation $\partial \dot{\boldsymbol{E}}_q' / \partial \boldsymbol{E}_q'$ is obtained as:

$$\frac{\partial \dot{\boldsymbol{E}}_q'}{\partial \boldsymbol{E}_q'} = \boldsymbol{T}_{d0}'^{-1}\left[2\mathrm{real}\left(\frac{\partial \boldsymbol{F}_{Eq}}{\partial \boldsymbol{V}}\frac{\partial \boldsymbol{V}}{\partial \boldsymbol{E}_q'}\right)+\frac{\partial \boldsymbol{F}_{Eq}}{\partial \boldsymbol{E}_q'}\right] \quad (24)$$

Then it can be deduced that the off-diagonal elements of (24) are non-negative numbers, which satisfies the monotone system condition to be described in next section, indicating that the SGs have cooperative voltage regulation property.

*2) The Partial Derivative Matrix* $\partial \dot{\boldsymbol{E}}_q' / \partial \boldsymbol{E}_{fd}$

From the interface equation of SGs, it follows that:

$$\frac{d\boldsymbol{V}}{d\boldsymbol{E}_{fd}} = \boldsymbol{O} \quad (25)$$

Therefore, one can directly obtain:

$$\frac{\partial \dot{\boldsymbol{E}}_q'}{\partial \boldsymbol{E}_{fd}} = \boldsymbol{T}_{d0}'^{-1} \quad (26)$$

which is a nonnegative diagonal matrix obviously.

*3) The Partial Derivative Matrix* $\partial \dot{\boldsymbol{E}}_{\mathrm{fd}} / \partial \boldsymbol{E}_{\mathrm{fd}}$

Similarly, one can have:

$$\frac{\partial \dot{\boldsymbol{E}}_{fd}}{\partial \boldsymbol{E}_{fd}} = -\boldsymbol{T}_{\mathrm{A}}^{-1} \quad (27)$$

It indicates that the excitation system satisfies the monotone system condition to be described in next section.

*4) The Partial Derivative Matrix* $\partial \dot{\boldsymbol{E}}_{fd} / \partial \boldsymbol{E}_q'$

The partial derivatives of the excitation system with respect to the voltage are:

$$\frac{\partial \boldsymbol{F}_{Efd}}{\partial \boldsymbol{V}} = \mathrm{conj}\left(\frac{\partial \boldsymbol{F}_{Efd}}{\partial \bar{\boldsymbol{V}}}\right) = \mathrm{diag}\left(-\frac{K_{Ai}\bar{V}_{ti}}{2\left|V_{ti}\right|}\right), \frac{\partial \boldsymbol{F}_{Efd}}{\partial \boldsymbol{E}_q'} = \boldsymbol{0} \quad (28)$$

Substituting (28) into (12) to get:

$$\frac{\partial \dot{\boldsymbol{E}}_{fd}}{\partial \boldsymbol{E}_q'} = 2\boldsymbol{T}_{\mathrm{A}}^{-1}\mathrm{real}\left(\frac{\partial \boldsymbol{F}_{Efd}}{\partial \boldsymbol{V}}\frac{\partial \boldsymbol{V}}{\partial \boldsymbol{E}_q'}\right) \quad (29)$$

It indicates that the Jacobian matrix of the excitation voltage with respect to the voltage behind transient reactance is a negative matrix, so the voltage behind transient reactance appears to be a negative feedback signal to the excitation system.

In summary, when all the generators in the power system are SGs, the Jacobian matrix of the system transient voltage subsystem has the sign pattern shown in Fig. 3. This result in general agrees with our previous findings [17], albeit it extends to more realistic power system models.

$$\mathrm{sgn}\left(\begin{bmatrix} \frac{\partial \dot{\boldsymbol{E}}_{fd}}{\partial \boldsymbol{E}_{fd}} & \frac{\partial \dot{\boldsymbol{E}}_{fd}}{\partial \boldsymbol{E}_q'} \\ \frac{\partial \dot{\boldsymbol{E}}_q'}{\partial \boldsymbol{E}_{fd}} & \frac{\partial \dot{\boldsymbol{E}}_q'}{\partial \boldsymbol{E}_q'} \end{bmatrix}\right) = \begin{bmatrix} \begin{bmatrix} - & 0 & 0 \\ 0 & - & 0 \\ 0 & 0 & - \end{bmatrix} & \begin{bmatrix} - & - & - \\ - & - & - \\ - & - & - \end{bmatrix} \\ \begin{bmatrix} + & 0 & 0 \\ 0 & + & 0 \\ 0 & 0 & + \end{bmatrix} & \begin{bmatrix} - & + & + \\ + & - & + \\ + & + & - \end{bmatrix} \end{bmatrix}$$

Fig. 3. Sign pattern of voltage subsystems of SGs only

*B. Sign Pattern of GFM Converters Voltage Dynamics*

According to the model (7) of the GFM converters, the partial derivatives of the GFM converter dynamic equations are as follows:

$$\begin{cases} \frac{\partial \boldsymbol{F}_{vir}}{\partial \boldsymbol{V}} = \mathrm{conj}\left(\frac{\partial \boldsymbol{F}_{vir}}{\partial \bar{\boldsymbol{V}}}\right) = \mathrm{diag}(K_{qi}x_{li}^{-1}(\bar{V}_{ti} - 0.5E_{viri}e^{-j\delta_i})) \\ \frac{\partial \boldsymbol{F}_{vir}}{\partial \boldsymbol{E}_{vir}} = \mathrm{diag}(-0.5K_{qi}x_{li}^{-1}(V_{ti}e^{-j\delta_i} + \bar{V}_{ti}e^{j\delta_i})) \\ \frac{\partial \boldsymbol{F}_{vir_fd}}{\partial \boldsymbol{V}} = \mathrm{conj}\left(\frac{\partial \boldsymbol{F}_{vir_fd}}{\partial \bar{\boldsymbol{V}}_{\mathrm{t}}}\right) = \mathrm{diag}\left(-\frac{K_{ui}\bar{V}_{ti}}{2\left|V_{ti}\right|}\right) \\ \frac{\partial \boldsymbol{F}_{vir_fd}}{\partial \boldsymbol{E}_{vir}} = \boldsymbol{0}, \frac{\partial \boldsymbol{F}_{vir_fd}}{\partial \boldsymbol{E}_{vir_fd}} = -\boldsymbol{I}, \frac{\partial \boldsymbol{F}_{vir}}{\partial \boldsymbol{E}_{vir_fd}} = \boldsymbol{I} \end{cases} \quad (30)$$

where $\mathrm{conj}(\cdot)$ denotes the usual conjugation operation. The above formulas show that the partial derivatives associated with GFM converter dynamics have the same sign pattern as those with SGs.

According to output power expression (8) for GFM converters, we get:

$$\begin{cases} \frac{\partial \bar{g}_{GFM,i}}{\partial V_i} = jx_{l,i}^{-1}\bar{V}_{ti} \\ \frac{\partial \bar{g}_{GFM,i}}{\partial \bar{V}_i} = -jE_{vir,i}x_{l,i}^{-1}e^{j\delta_i} + jx_{l,i}^{-1}V_{ti} \\ \frac{\partial \bar{g}_{GFM,i}}{\partial E_{vir,i}} = (\sin\delta_i - j\cos\delta_i)\bar{V}_{ti}x_{l,i}^{-1} \end{cases} \quad (31)$$

It comes as no surprise that the sign pattern of Jacobian matrix of GFM converter dynamics is exactly the same as that of a SG.

So far, we have deduced the sign pattern of Jacobian matrix of the transient voltage subsystem. The findings will be later utilized to study transient voltage behavior of power systems with GFM converters. It should be noted that an actual GFM converter may experience control mode switching especially under low-voltage conditions, this is left as an interesting topic for future work. For convenience, let $\boldsymbol{\mathcal{E}}$ denote the voltages behind transient reactance of SGs and the internal voltages of GFM converters, and $\boldsymbol{\mathcal{E}}_{\mathrm{fd}}$ denote the excitation voltages of SGs and GFM converters.

## IV. Review of Monotone Dynamic Systems Theory

This section reviews the concepts of monotone system theory, which later prove to be useful. For more details on the theoretical developments, the readers are referred to [24]-[26].

*Definition 1*: For a nonlinear control system:

$$\Sigma_1 : \dot{\boldsymbol{x}} = \boldsymbol{f}(\boldsymbol{x}, \boldsymbol{v}) \quad (32)$$

if the following condition is met at $t \geq 0$:

$$\boldsymbol{v}_1 \succeq \boldsymbol{v}_2, \xi_1 \succeq \xi_2 \Rightarrow \phi(\xi_1) \succeq \phi(\xi_2) \quad (33)$$

where $\boldsymbol{x} \succeq \boldsymbol{y}$ means $\boldsymbol{x}_i \geq \boldsymbol{y}_i$ holds for all $i$, then we call $\Sigma_1$ an input-state monotone system.

In this and subsequent sections, $\boldsymbol{x} \in R^n$ is the state variable vector, $\boldsymbol{v} \in R^p$ is the input variable vector, $\boldsymbol{f}$ is locally Lipschitz continuous in $\boldsymbol{x}$ and jointly continuous in $\boldsymbol{x}$ and $\boldsymbol{v}$, and $\phi(\xi)$ is the solutions satisfying $\boldsymbol{x}(0) = \xi$.

*Definition 2*: For a monotone nonlinear controlled system with outputs:

$$\Sigma_2 : \dot{\boldsymbol{x}} = \boldsymbol{f}(\boldsymbol{x},\boldsymbol{v}),\ \boldsymbol{y} = \boldsymbol{h}(\boldsymbol{x}) \tag{34}$$

if the output function $\boldsymbol{h}$ satisfies the following condition:

$$\boldsymbol{v}_1 \succeq \boldsymbol{v}_2, \xi_1 \succeq \xi_2 \Rightarrow \phi(\xi_1) \succeq \phi(\xi_2), \boldsymbol{h}(\xi_1) \succeq \boldsymbol{h}(\xi_2) \tag{35}$$

then the system $\Sigma_2$ is called an input-state-output monotone system.

The input-state-output monotonicity of the systems can be determined based on the sign pattern of the trajectory Jacobian matrix.

*Theorem 1 [27]*: Nonlinear controlled system $\Sigma_2$ is input-state-output monotone if

$$\begin{cases} \dfrac{\partial f_i}{\partial x_j}(x,v) \geq 0 & \forall i \neq j \\ \dfrac{\partial f_i}{\partial v_j}(x,v) \geq 0 & \forall i,j \\ \dfrac{\partial h_i}{\partial x_j}(x) \geq 0 & \forall i,j \end{cases} \tag{36}$$

Below a simple linear system is used to illustrate the above result. The system model is as follows

$$\begin{bmatrix} \dot{x}_1 \\ \dot{x}_2 \\ \dot{x}_3 \end{bmatrix} = \begin{bmatrix} -1 & 0 & 2 \\ 1 & -3 & 0 \\ 0 & 1 & -4 \end{bmatrix} \begin{bmatrix} x_1 \\ x_2 \\ x_3 \end{bmatrix} + \begin{bmatrix} 4 \\ 0 \\ 1 \end{bmatrix} v$$
$$y = \begin{bmatrix} 0 & 1 & 2 \end{bmatrix} \begin{bmatrix} x_1 & x_2 & x_3 \end{bmatrix}^T \tag{37}$$

Obviously, the system satisfies the conditions of Theorem 1. Assuming that the initial values of the state variables are $\boldsymbol{x}$ = [0,0,0], the response curves of the state variable and the output signal are observed under the input signals with unit step response $u(t)$ and $2u(t)$, and the results are shown below.

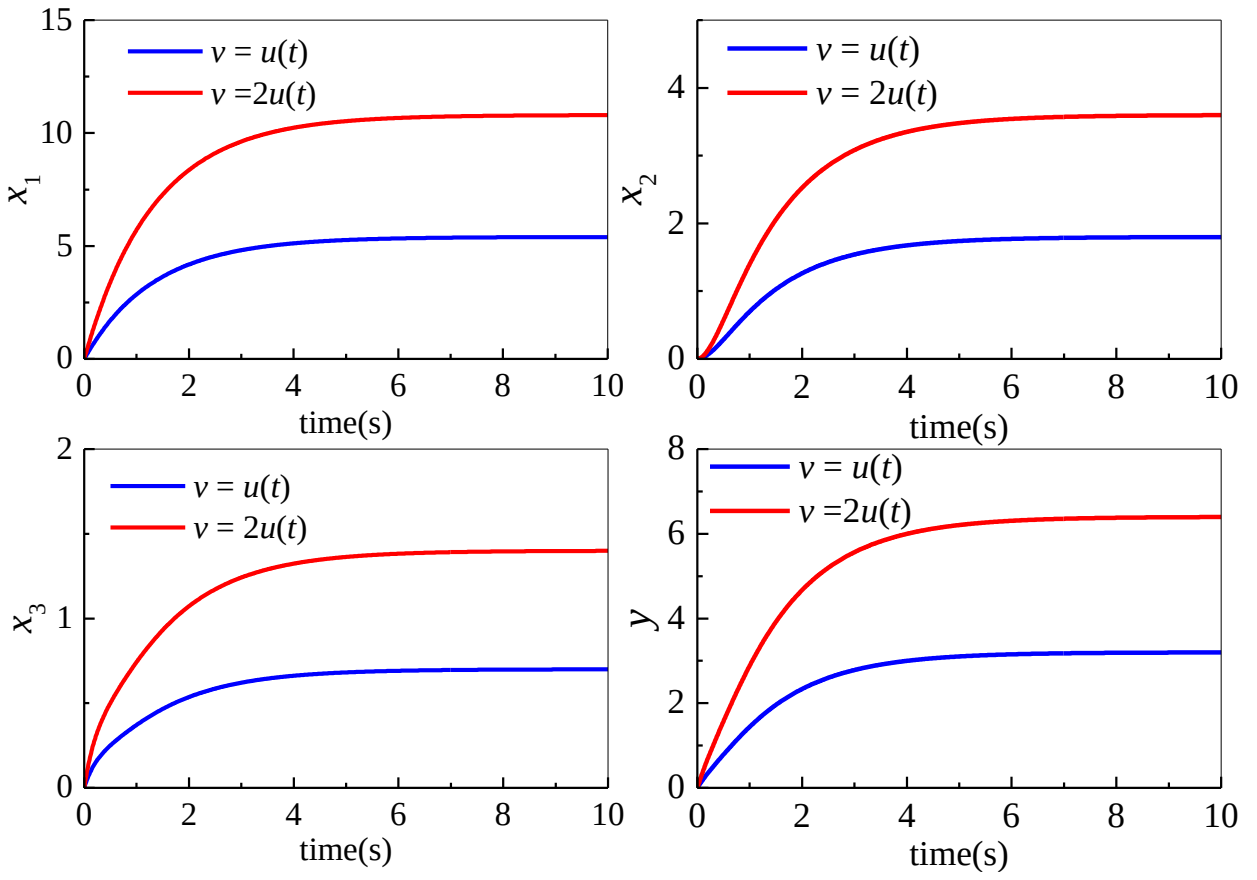

Fig. 4. Time-domain response solutions for the linear system

From Fig. 4, it can be observed that all state variables and output signal consistently maintain the input-state-output monotonicity relationship with the input signal.

Quite often the voltage dynamics of power systems may not always meet the conditions of the above result, however, under such occasions a weaker form of monotonicity, that is, input-output monotonicity holds true. Input-output monotonicity implies that the input and output maintain the relationship defined in (35) even though the solutions of the states do not necessarily adhere to it. Section V describes how to apply *Theorem 1* to study such input-output monotonicity of voltage trajectories.

## V. Input-Output Monotonicity of Voltage Trajectories

This section examines the input-output monotonicity of voltage trajectories of power systems with GFM converters.

### *A. A Reduced-Order Model for Studying Voltage Behavior*

To gain better understanding into voltage dynamics of power systems with GFM converters, equations (5) and (7) can be reduced based on singular perturbation theory. It is not surprising that the voltage subsystem has two time-scales, in which the internal potential system is a slow system while the excitation system is a fast system.

The celebrated Tihkonov's theorem states that [28]: if the following two assumptions hold, 1) the equilibrium of the fast system is asymptotically stable and the initial states are in the stability region; 2) the real parts of the eigenvalues of $\partial\dot{\mathcal{E}}_{fd}/\partial\mathcal{E}_{fd}$ along the trajectories are negative; then the dynamics of $\mathcal{E}_{fd}$ will quickly converge to the quasi-steady state determined by the algebraic manifold $\boldsymbol{g} = 0$.

As has been validated in Section II, the Jacobian matrix of the excitation system is

$$\frac{\partial\dot{\mathcal{E}}_{fd}}{\partial\mathcal{E}_{fd}} = -\boldsymbol{T}_{Efd}^{-1} \tag{38}$$

where $\boldsymbol{T}_{\mathrm{E}fd}$ is the time constant. As the real parts of the eigenvalues of $\partial\dot{\mathcal{E}}_{fd}/\partial\mathcal{E}_{fd}$ are always negative and the excitation system is decoupled, assumption 1 and 2 both hold true.

The above arguments show that the dynamics of the fast system can be effectively reduced to manifold $\boldsymbol{g} = 0$. Taking SG as an example, that is

$$\boldsymbol{g} = \boldsymbol{K}_A(\boldsymbol{V}_{ref} - |\boldsymbol{V}_t|) - \boldsymbol{E}_{fd} = \boldsymbol{0} \tag{39}$$

Substituting the excitation voltage expression into the dynamic equation of the voltages behind transient reactance yields the reduced-order model

$$\boldsymbol{T}'_{d0}\dot{\boldsymbol{E}}'_q = (\boldsymbol{x}_d \boldsymbol{x}'^{-1}_d - \boldsymbol{I})\boldsymbol{V}_q + \boldsymbol{K}_A(\boldsymbol{V}_{ref} - |\boldsymbol{V}_t|) - \boldsymbol{x}_d \boldsymbol{x}'^{-1}_d \boldsymbol{E}'_q \tag{40}$$

Similarly, the reduced-order model of GFM converters can also be obtained as

$$\boldsymbol{K}_i \dot{\boldsymbol{E}}_{vir} = \boldsymbol{K}_u(\boldsymbol{V}_{ref} - |\boldsymbol{V}_t|) + \boldsymbol{K}_q(\boldsymbol{Q}_{ref} - \boldsymbol{Q}_e) \tag{41}$$

The Jacobian matrix $\boldsymbol{J}$ of the reduced-order model can be obtained from the aforementioned partial derivation matrix

$$\begin{aligned} \boldsymbol{J} &= \frac{\partial\dot{\mathcal{E}}}{\partial\mathcal{E}} + \boldsymbol{T}_{\mathrm{E}fd}\boldsymbol{T}_E^{-1}\frac{\partial\dot{\mathcal{E}}_{\mathrm{fd}}}{\partial\mathcal{E}} \\ &= \boldsymbol{T}_E^{-1}\left[2\mathrm{real}\left(\frac{\partial\boldsymbol{F}_E}{\partial\boldsymbol{V}}\frac{\partial\boldsymbol{V}}{\partial\mathcal{E}}\right) + \frac{\partial\boldsymbol{F}_E}{\partial\mathcal{E}} + 2\boldsymbol{T}_{\mathrm{E}fd}\mathrm{real}\left(\frac{\partial\boldsymbol{F}_{\mathrm{E}fd}}{\partial\boldsymbol{V}}\frac{\partial\boldsymbol{V}}{\partial\mathcal{E}}\right)\right] \end{aligned} \tag{42}$$

where $\boldsymbol{T}_{\mathrm{E}}$ is the time constant. It is obvious that $\partial\boldsymbol{F}_{\mathrm{E}}/\partial\mathcal{E}$, $\partial\boldsymbol{F}_{\mathrm{E}}/\partial\boldsymbol{V}$ and $\partial\boldsymbol{F}_{\mathrm{Efd}}/\partial\boldsymbol{V}$ are diagonal matrixes because the dynamic equations of the power components contain only their own variables. According to the diagonal dominance property of impedance matrix $\boldsymbol{Z}$, it can be deduced that $\partial \boldsymbol{V}/\partial \boldsymbol{\mathcal{E}}$ is also a diagonally dominant matrix. Therefore, the Jacobian matrix $\boldsymbol{J}$ has approximately diagonally dominant property, and the diagonal elements are all negative.

The sign pattern of the non-diagonal elements of the Jacobian matrix $\boldsymbol{J}$ is affected by the magnitude of exciter gains which can be deduced from (28) and (30). Take (28) for example,

$$\frac{\partial \boldsymbol{F}_{Efd}}{\partial \boldsymbol{V}} = \text{diag}\left(-\frac{K_{Ai}\bar{V}_i}{2T_{Ai}V_{ti}}\right) \tag{43}$$

when the excitation gains are relatively low, because the non-diagonal elements of

$$\text{real}\left(\frac{\partial \boldsymbol{F}_E}{\partial \boldsymbol{V}} \frac{\partial \boldsymbol{V}}{\partial \boldsymbol{\mathcal{E}}}\right) \tag{44}$$

is positive, so the non-diagonal elements of the Jacobian matrix $\boldsymbol{J}$ are also positive, in which case the reduced-order system is a cooperative system. Conversely, the reduced-order system is a competitive system, i.e., the non-diagonal elements are negative when the excitation gains are relatively high.

For voltage trajectories, it can be obtained that

$$\frac{\partial |V_i|}{\partial \mathcal{E}_j} = \frac{1}{|V_i|}\text{real}(V_i \frac{\partial \bar{V}_i}{\partial \mathcal{E}_j}) \tag{45}$$

Based on the above analysis, it is easy to conclude that $\partial|\boldsymbol{V}|/\partial\boldsymbol{\mathcal{E}}$ is a non-negative matrix (see Fig. 5).

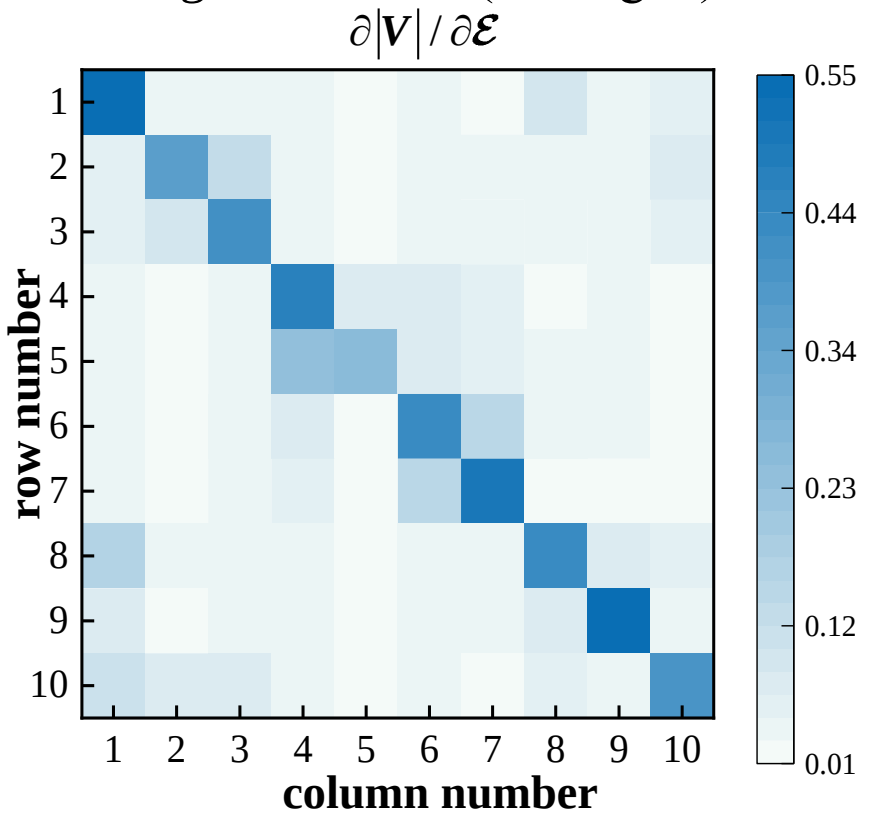

Fig. 5. The partial derivative matrix of 10-machine 39-bus test system

It can be seen from Fig. 5 that, the partial derivative matrix $\partial|\boldsymbol{V}|/\partial\boldsymbol{\mathcal{E}}$ is non-negative, it is also diagonal dominant.

*B. Input-State-Output Monotonicity under Low Excitation Gains*

Power electronic equipment is characterized by flexible control and fast regulation capabilities, so in the modern power system it is effective to regulate voltage by adjusting parameters of a GFM converter. In this section, our interest is to study the effect of adjusting reference voltage $V_{ref}$ (see Fig. 1) on dynamic voltage behavior.

As shown in previous section, it is typical that the voltage subsystem can be either cooperative or competitive. Let us assume that the excitation gains are relatively low, so the non-diagonal elements of the Jacobian matrix $\boldsymbol{J}$ are all positive [29]. Let $\boldsymbol{v} = \boldsymbol{V}_{ref}$ and $\boldsymbol{y} = |\boldsymbol{V}|$ be network voltage vector, further let $\boldsymbol{x} = \boldsymbol{\mathcal{E}}$, the voltage subsystems (40) and (41) can be re-written in compact form as:

$$\Sigma_2 : \dot{\boldsymbol{x}} = \boldsymbol{f}(\boldsymbol{x},\boldsymbol{v}),\ \boldsymbol{y} = \boldsymbol{h}(\boldsymbol{x}) \tag{46}$$

It follows immediately from *Theorem 1* that the voltage trajectories under reference voltage control exhibits input-state-output monotonicity property. To see how this works, let $\Delta\boldsymbol{x}$, $\Delta\boldsymbol{y}$ be the corresponding incremental vector of $\boldsymbol{x}$, $\boldsymbol{y}$. Consider the variation equation [30],

$$\Delta\dot{\boldsymbol{x}} = \underbrace{\frac{\partial \boldsymbol{f}}{\partial \boldsymbol{x}}}_{\boldsymbol{\Gamma}}\Delta\boldsymbol{x} + \underbrace{\frac{\partial \boldsymbol{f}}{\partial \boldsymbol{v}}}_{\boldsymbol{\Lambda}}\Delta\boldsymbol{v} \tag{47}$$

Assume $\Delta\boldsymbol{x}(0) \geq 0$, $\Delta\boldsymbol{v}(0) \geq 0$, now suppose there exists $\tau > 0$ such that $\Delta\boldsymbol{x}_i(\tau) = 0$, $\Delta\boldsymbol{x}_j(\tau) \geq 0$ $(j \neq i)$, it follows that

$$\begin{aligned}\Delta\dot{\boldsymbol{x}}_i(\tau) &= \sum_{j=1}^{n}\boldsymbol{\Gamma}_{ij}(\tau)\Delta\boldsymbol{x}_j(\tau) + \sum_{j=1}^{m}\boldsymbol{\Lambda}_{ij}(\tau)\Delta\boldsymbol{u}_j(\tau) \\ &= \sum_{j=1, j\neq i}^{n}\boldsymbol{\Gamma}_{ij}(\tau)\Delta\boldsymbol{x}_j(\tau) + \sum_{j=1}^{m}\boldsymbol{\Lambda}_{ij}(\tau)\Delta\boldsymbol{u}_j(\tau) \geq 0\end{aligned} \tag{48}$$

This shows that $\Delta\boldsymbol{x}(t) \geq 0$, i.e., the variation system is positive. Fig. 6 illustrates the trajectory of a positive variation system. It can be observed that the trajectory originating in first quadrant will remain within this quadrant forever, because as soon as the trajectory touches the boundary of the first quadrant, it bounces off.

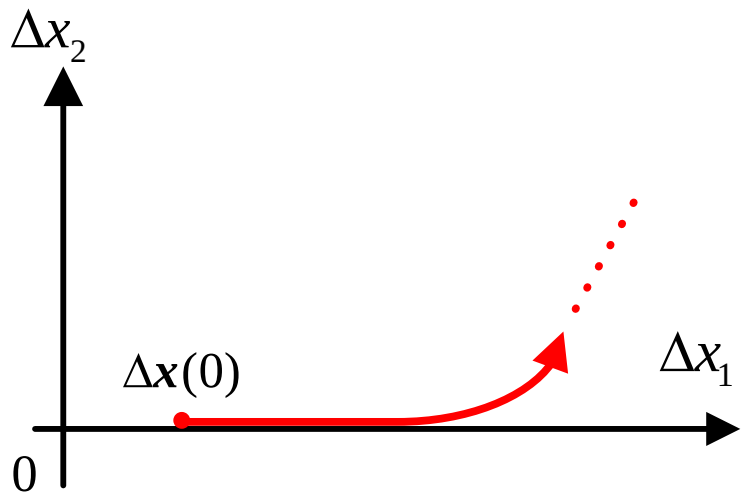

Fig. 6. The trajectory of a cooperative system

In addition, since $\partial\boldsymbol{h}/\partial\boldsymbol{x}$ is a positive matrix, we have $\Delta\boldsymbol{y}(t) \geq 0$. This establishes that the voltage subsystem $\Sigma_2$ is locally input-state-output monotone. By invoking mean value theorem [27], it is not difficult to see that, remarkably, the input-state-output monotonicity property of $\Sigma_2$ holds globally.

*C. Input-State-Output Monotonicity under High Excitation Gains*

Now assume that the excitation gains are relatively high, in which case the voltage subsystem becomes competitive. Apparently, the voltage subsystem can still be re-written in the following compact form:

$$\Sigma_2 : \dot{\boldsymbol{x}} = \boldsymbol{f}(\boldsymbol{x},\boldsymbol{v}),\ \boldsymbol{y} = \boldsymbol{h}(\boldsymbol{x})$$

with $\boldsymbol{v} = V_{\text{ref}}$, $\boldsymbol{y} = |\boldsymbol{V}|$ and $\boldsymbol{x} = \boldsymbol{\mathcal{E}}$.

In Case2 mentioned in Section VI, the values of Jacobian matrix $\boldsymbol{J}$ were calculated and the results are shown in Fig. 7.

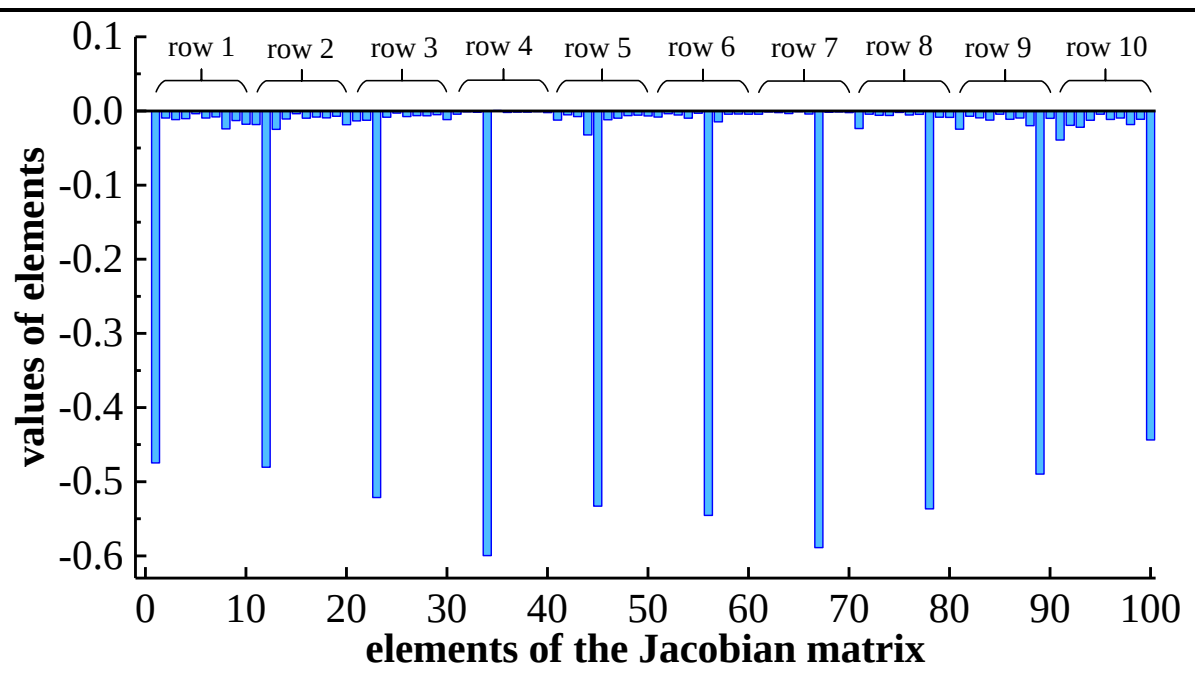

Fig. 7. The values of the elements of the Jacobian matrix of the reduced-order system in Case 2

It is obvious that the non-diagonal elements are much smaller than diagonal elements, which is consistent with the analysis. By the way, according to Gersgorin's disc theorem [31], the eigenvalues of the Jacobian matrix $\boldsymbol{J}$ all lie in the left half-plane, and the absolute values of the real parts are much larger than the imaginary parts. This means that the reduced-order system usually possesses very high stability margin.

Back to GFM control, in (41), it is obvious that

$$\frac{\partial \dot{\boldsymbol{E}}_{\text{vir}}}{\partial \boldsymbol{V}_{\text{ref}}} = \frac{\boldsymbol{K}_{\text{u}}}{\boldsymbol{K}_{\text{i}}} \geq 0 \quad \text{and} \quad \frac{\partial \dot{\boldsymbol{E}}_{\text{vir}}}{\partial \boldsymbol{Q}_{\text{ref}}} = \frac{\boldsymbol{K}_{\text{q}}}{\boldsymbol{K}_{\text{i}}} \geq 0 \tag{49}$$

Now consider again the expression of $\Delta \dot{x}_i(\tau)$, we claim that:

$$\begin{aligned} \Delta \dot{\boldsymbol{x}}_i(\tau) &= \sum_{j=1, j\neq i}^{n} \boldsymbol{\Gamma}_{ij}(\tau)\Delta \boldsymbol{x}_j(\tau) + \sum_{j=1}^{m} \boldsymbol{\Lambda}_{ij}(\tau)\Delta \boldsymbol{u}_j(\tau) \\ &\approx \sum_{j=1}^{m} \boldsymbol{\Lambda}_{ij}(\tau)\Delta \boldsymbol{u}_j(\tau) \\ &\geq 0 \end{aligned} \tag{50}$$

The reason is that, while the first term in the inequality now becomes negative, however, its magnitude is much smaller than the second term which is positive. This demonstrates that $\Delta \boldsymbol{x}(t) \geq 0$ (see Section VI.B for an example, especially (57)).

The above results show that, when adjusting the reference values of voltage or reactive power of a GFM converter, the network voltage trajectories vary in a consistent way (see Fig. 12 in section VI). This fact can be utilized to guide the design of future advanced GFM controls (refer to [32] for instance).

*D. Input-Output Monotonicity under Load-shedding Controls*

It is not difficult to understand that load-shedding operations enjoy similar input-output monotonicity revealed in previous section. However, a rigorous analysis for this well-known correct control strategy is not available in the literature. This section provides an interpretation for this phenomenon. The voltage subsystem model for load-shedding control is somewhat different from (34) as shown below:

$$\Sigma_3 : \dot{\boldsymbol{x}} = \boldsymbol{f}(\boldsymbol{x}, \boldsymbol{v}),\ \boldsymbol{y} = \boldsymbol{h}(\boldsymbol{x}, \boldsymbol{v}) \tag{51}$$

here the input $\boldsymbol{v}$ denotes the amount of load-shedding, while $\boldsymbol{y} = |\boldsymbol{V}|$ and $\boldsymbol{x} = \boldsymbol{\mathcal{E}}$. System $\Sigma_3$ does not possess input-state-output monotonicity property, as $\partial \boldsymbol{f}/\partial \boldsymbol{x}$ is not a cooperative matrix. Nevertheless, it still possesses a weaker form of monotonicity, namely, input-output monotonicity.

Suppose $\boldsymbol{\phi}(t, \boldsymbol{x}_0, \boldsymbol{v})$ and $\boldsymbol{y}(t, \boldsymbol{\phi}(t, \boldsymbol{x}_0, \boldsymbol{v}), \boldsymbol{v})$ denote the solutions of $\boldsymbol{x}$ and $\boldsymbol{y}$, respectively. For $\boldsymbol{v}_1 \succeq \boldsymbol{v}_2$ and $t \geq 0$, the input-output monotonicity implies that

$$\Delta \boldsymbol{y} = \boldsymbol{y}(t, \boldsymbol{\phi}_1(t, \boldsymbol{x}_0, \boldsymbol{v}_1), \boldsymbol{v}_1) - \boldsymbol{y}(t, \boldsymbol{\phi}_2(t, \boldsymbol{x}_0, \boldsymbol{v}_2), \boldsymbol{v}_2) \geq 0 \tag{52}$$

For simplicity, we replace $\boldsymbol{\phi}(t, \boldsymbol{x}_0, \boldsymbol{v})$ with $\boldsymbol{\phi}$ in the following text. To establish the above result, firstly, it is necessary to define a function

$$\Upsilon(\sigma) = \boldsymbol{y}(t, \boldsymbol{\phi}_2 + \sigma(\boldsymbol{\phi}_1 - \boldsymbol{\phi}_2), \boldsymbol{v}_2 + \sigma(\boldsymbol{v}_1 - \boldsymbol{v}_2)) \tag{53}$$

Then

$$\boldsymbol{y}(t, \boldsymbol{\phi}_1, \boldsymbol{v}_1) - \boldsymbol{y}(t, \boldsymbol{\phi}_2, \boldsymbol{v}_2) = \Upsilon(1) - \Upsilon(0) = \int_0^1 \Upsilon'(\sigma) d\sigma \tag{54}$$

where

$$\Upsilon'(\sigma) = \frac{d\Upsilon}{d\sigma} = \frac{\partial \boldsymbol{h}}{\partial \boldsymbol{x}}(\boldsymbol{\phi}_1 - \boldsymbol{\phi}_2) + \frac{\partial \boldsymbol{h}}{\partial \boldsymbol{v}}(\boldsymbol{v}_1 - \boldsymbol{v}_2) \tag{55}$$

Hence, in order to determine the input-output monotonicity of the voltage trajectories, it suffices to show $\Upsilon'(\sigma) \geq 0$ for $t \geq 0$.

To proceed, let us see an example of 575-machine, 3243-node, 8117-bus East China power grid (a simplified system from the actual East China power grid) [33]. The initial voltage of bus WanBoYang was 0.9793 p.u and load shedding $\Delta \boldsymbol{Q}$ happened at $t = 0.5$s. The relationship between voltage of bus WanBoYang and load shedding MW at zone WH is shown in Fig. 8.

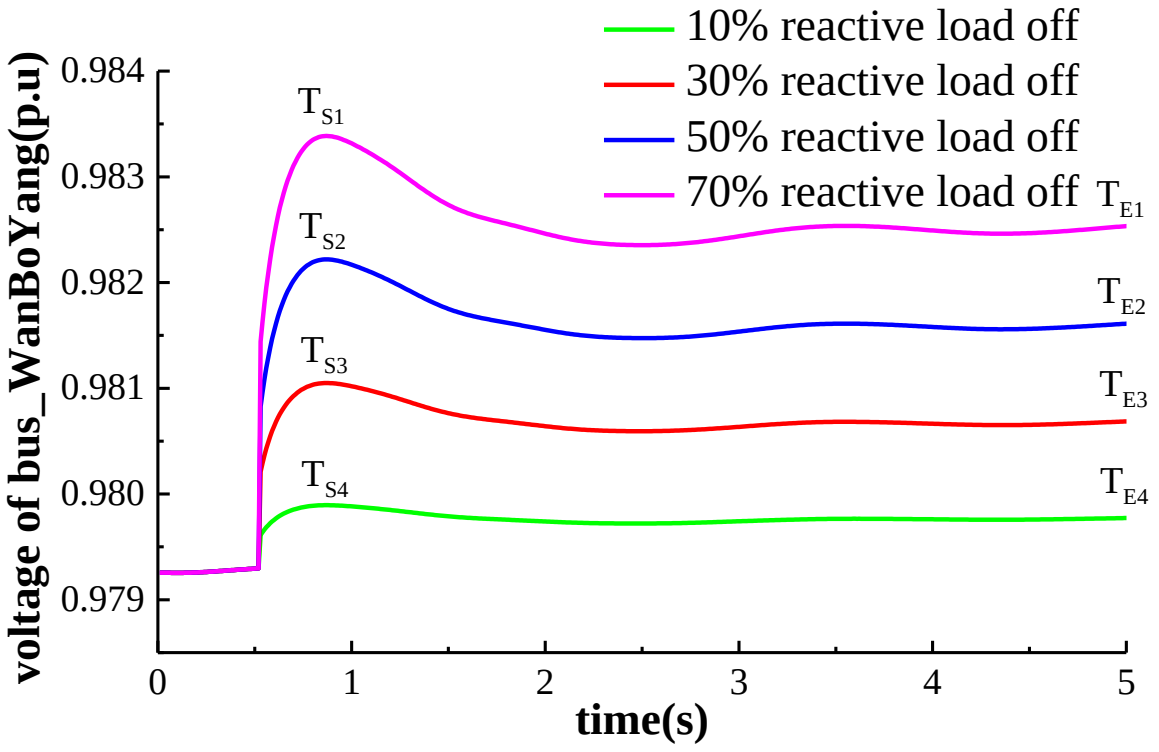

Fig. 8. The relationship between voltage variation and load shedding

It is obvious that during the short period of time after load shedding (near $T_{S1}$, $T_{S2}$, $T_{S3}$ and $T_{S4}$), the state variables cannot change significantly because the time constants of $\boldsymbol{\mathcal{E}}$ are relatively long, so $\boldsymbol{\mathcal{E}}_1 - \boldsymbol{\mathcal{E}}_2 \approx 0$, notice that $\boldsymbol{y} = |\boldsymbol{V}|$, it follows

$$\begin{aligned} \Upsilon'(\sigma) &= \frac{d|\boldsymbol{V}|(\sigma)}{d\sigma} \\ &= \frac{\partial |\boldsymbol{V}|}{\partial \boldsymbol{\mathcal{E}}}(\boldsymbol{\mathcal{E}}_1 - \boldsymbol{\mathcal{E}}_2) + \frac{\partial |\boldsymbol{V}|}{\partial \boldsymbol{Q}}(\boldsymbol{Q}_1 - \boldsymbol{Q}_2) \\ &\approx \frac{\partial |\boldsymbol{V}|}{\partial \boldsymbol{Q}}(\boldsymbol{Q}_1 - \boldsymbol{Q}_2) \end{aligned} \tag{56}$$

It is empirically known that $\partial|\boldsymbol{V}|/\partial \boldsymbol{Q}$ is always greater than zero [29],[33], this shows that $\Upsilon'(\sigma) \geq 0$.

As the system approaches the equilibrium point (near $T_{E1}$-$T_{E4}$), the inequality still holds true, due to the fact that the effect of the load shedding on the $\boldsymbol{\mathcal{E}}$ is insignificant, i.e., the variations of $\Delta\boldsymbol{\mathcal{E}}$ are small. This implies that $\Upsilon'(\sigma) \geq 0$.

By the same token, during the transient period (between $T_{S1}$-$T_{E1}$,..., $T_{S4}$-$T_{E4}$), the state variables do not differ much from it at the equilibrium point, so the inequality still holds. Therefore, one is still able to conclude that $\Upsilon'(\sigma) \geq 0$.

We end this section by noting that, the suggested monotone system approach suits very well in explaining the input-output monotonicity of power system voltage dynamics, to the best of the authors' knowledge, this is the first time such voltage behavior is rigorously studied.

## VI. Simulation Results

Aside from the simulation results of East China Power System shown in Section IV, this section provides additional results of the 10-machine 39-bus system taken from [34] using a prototype electro-mechanical simulation platform. Specifically, we report the results of the following two cases.

Case1: The generators are all SGs.

Case2: Replacing No.1-4 SGs with GFM converters with parameters shown in TABLE I.

TABLE I
Data of GFM converters

| parameters | $K_i$ /s | $K_q$ | $T_u$/s | $K_u$ |
|---|---|---|---|---|
| values | 0.1 | 5 | 0.02 | 1 |

A single line diagram of the 10-machine 39-bus power system is provided in Fig. 9.

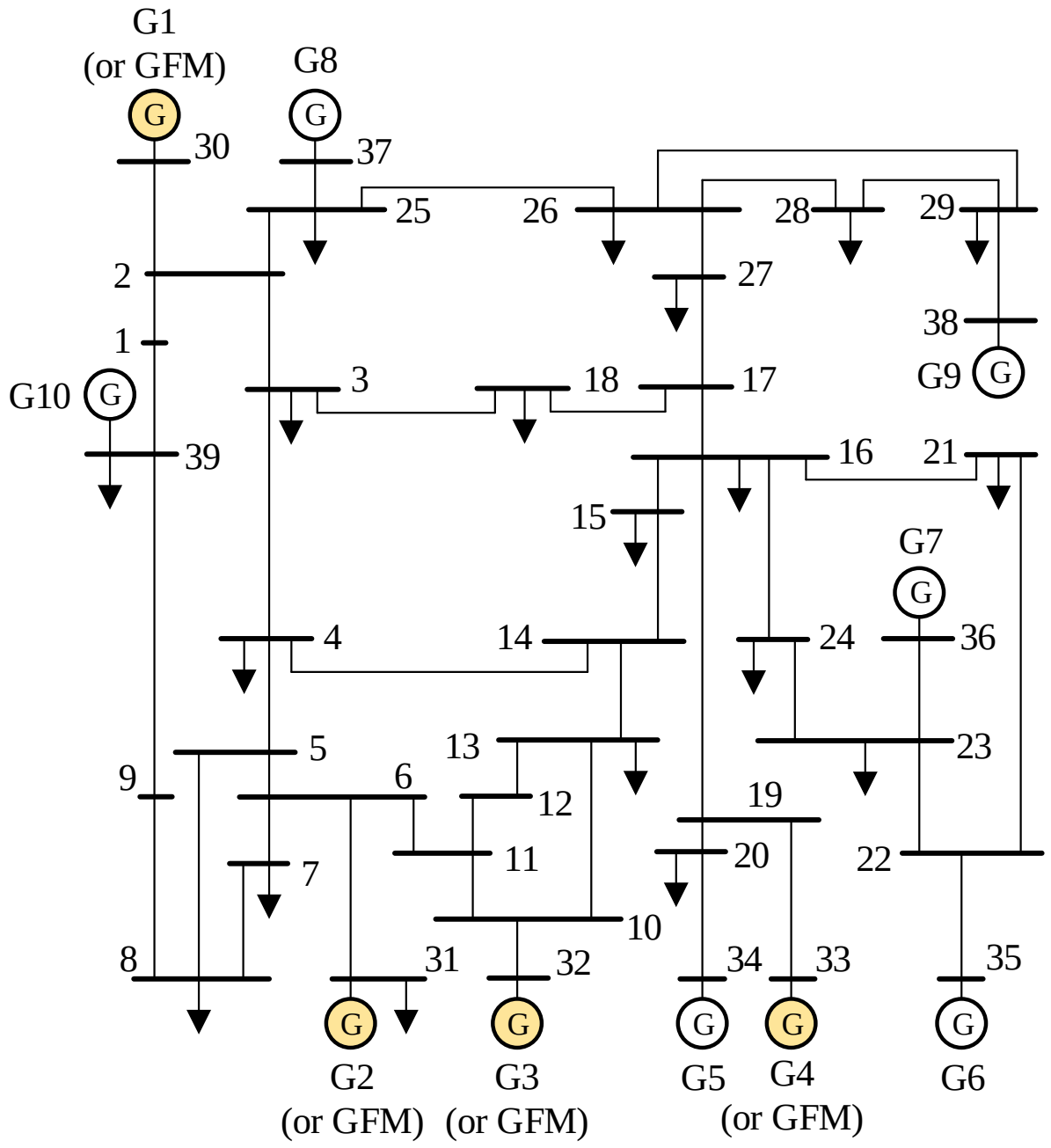

Fig. 9. The 10-machine 39-bus power system

Of particular interest here is the stable voltage dynamics, for results of unstable and critically stable systems, the readers are referred to (say) [33] and references cited therein.

### A. The Sign Pattern of Trajectory Jacobian Matrix

In this subsection, the sign pattern of trajectory Jacobian matrix indicated in Section II is verified, the condition when the sign pattern diminishes is discussed.

In Case1, a three-phase short-circuit fault is set at $t = 0.1$ s, and cleared after 0.06 s before the rotor angles swing significantly. For brevity, the curves of certain elements of Jacobian matrix are shown in Fig. 10.

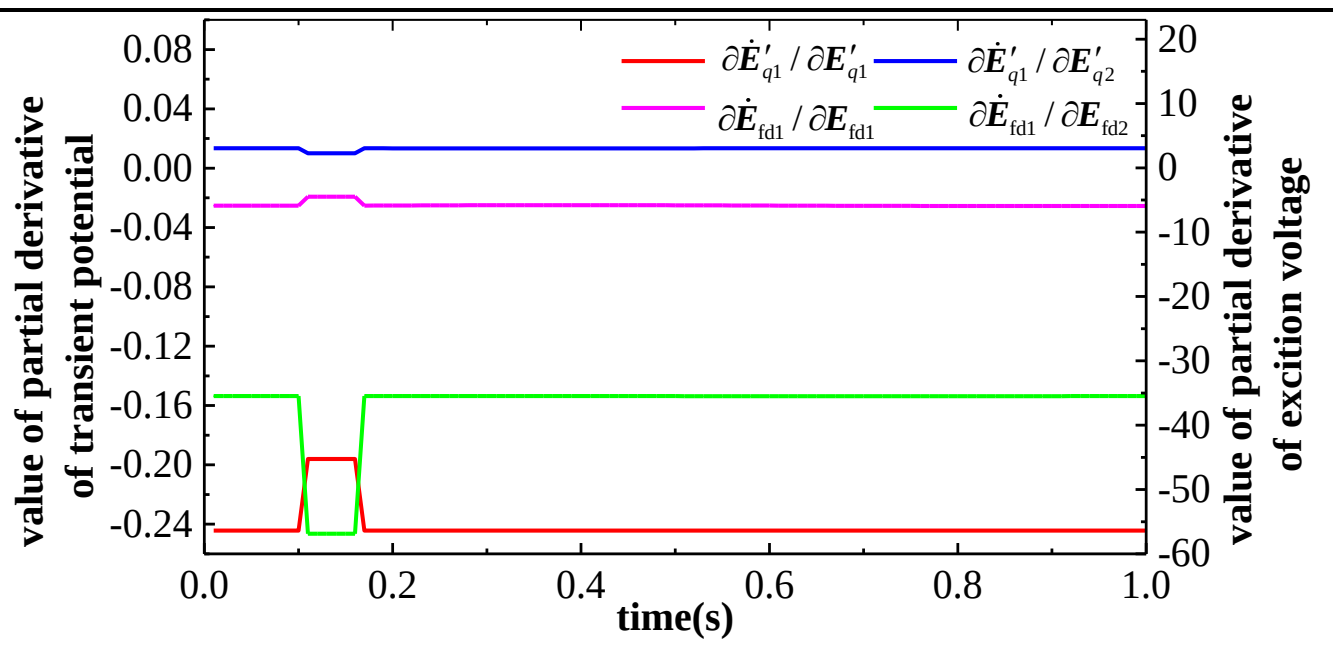

Fig. 10. Time-domain curves for nonzero elements of the Jacobian matrix when the rotor angles don't swing significantly

It can be seen from Fig. 10 that, the sign pattern of the Jacobian matrix are in accordance with the sign pattern of interconnected monotone system shown in Fig. 3 throughout the dynamic period. This validates that monotonicity is an intrinsic characteristics of the transient voltage subsystem.

It is not difficult to understand that, if rotor angles swing significantly, the monotonicity of the transient voltage subsystem would diminish. To observe that, the results of Case 1 are briefly described here. A three-phase short-circuit fault is set at $t = 0.1$ s, but cleared after 0.15 s. The curves of some elements of Jacobian matrix are shown in Fig. 11.

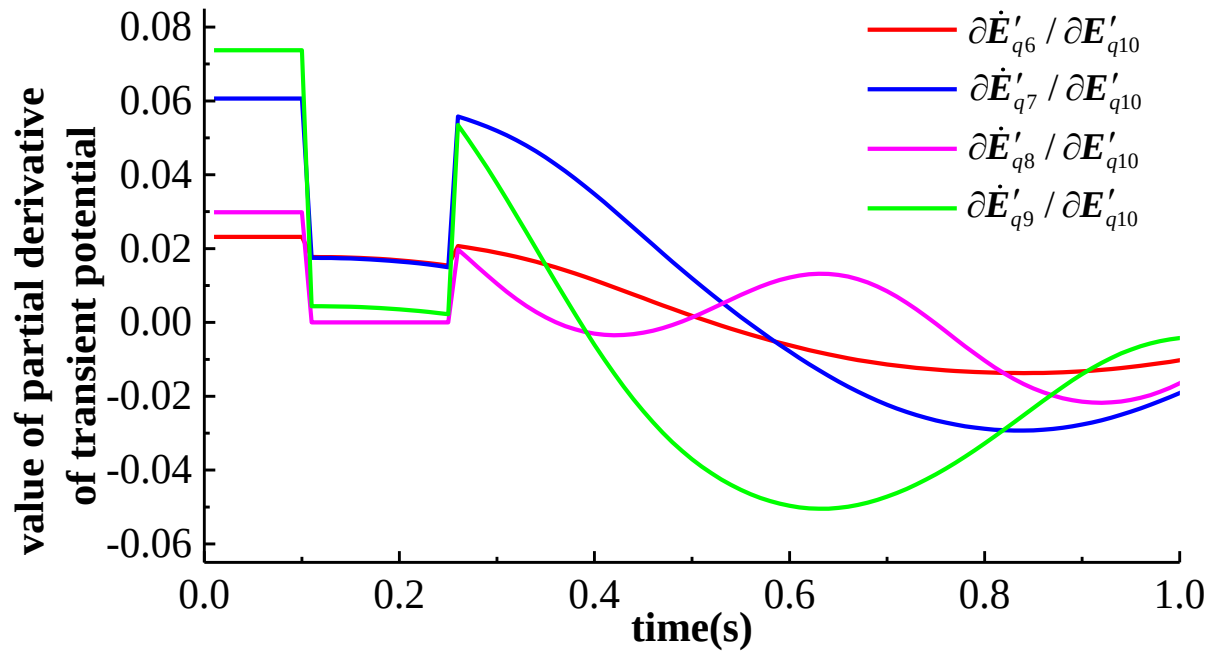

Fig. 11. Time-domain variation curves for nonzero elements of the Jacobian matrix when rotor angle swing significantly

It can be seen from Fig. 11 that, since rotor angles swing significantly, the sign pattern discussed in previous sections disappear. For such cases, the rotor angle stability and voltage stability of the power system are intertwined to affect the overall stability of the system. The complexity of such transient behavior (refer to, say, [35]) would require a more in-depth analysis.

The three-phase-to-ground fault is also set in Case2, and the sign change of Jacobian matrix during the whole dynamic process is in accordance with the sign pattern of Fig. 3. This proves that GFM converter can simulate the voltage regulation capability of SG and maintain the interconnected monotone system characteristics.

### B. Input-Output Monotonicity of Voltage Trajectories

In this subsection, the input-output monotonicity of the system voltage trajectories is validated.

#### 1) Voltage Trajectories under GFM Controls

With the parameters shown in TABLE I, it can be seen that

$$\Lambda_{11} = \frac{\partial \dot{E}_{vir,1}}{\partial V_{ref,1}} = \frac{K_u}{K_i} = 10 \tag{57}$$

It is significantly larger than the elements in Fig. 7, which implies that the input-output monotonicity of the voltage trajectories should hold true. In order to verify the input-output monotonicity of the voltage trajectories when the voltage reference of GFM converter varies, the voltage reference of the No. 1 GFM converter at bus30 was first raised at $t$ = 1 s and then decreased at $t$ = 3 s. The voltage trajectories of internal voltage of GFM converter and the voltage of bus30 are shown in Fig. 12.

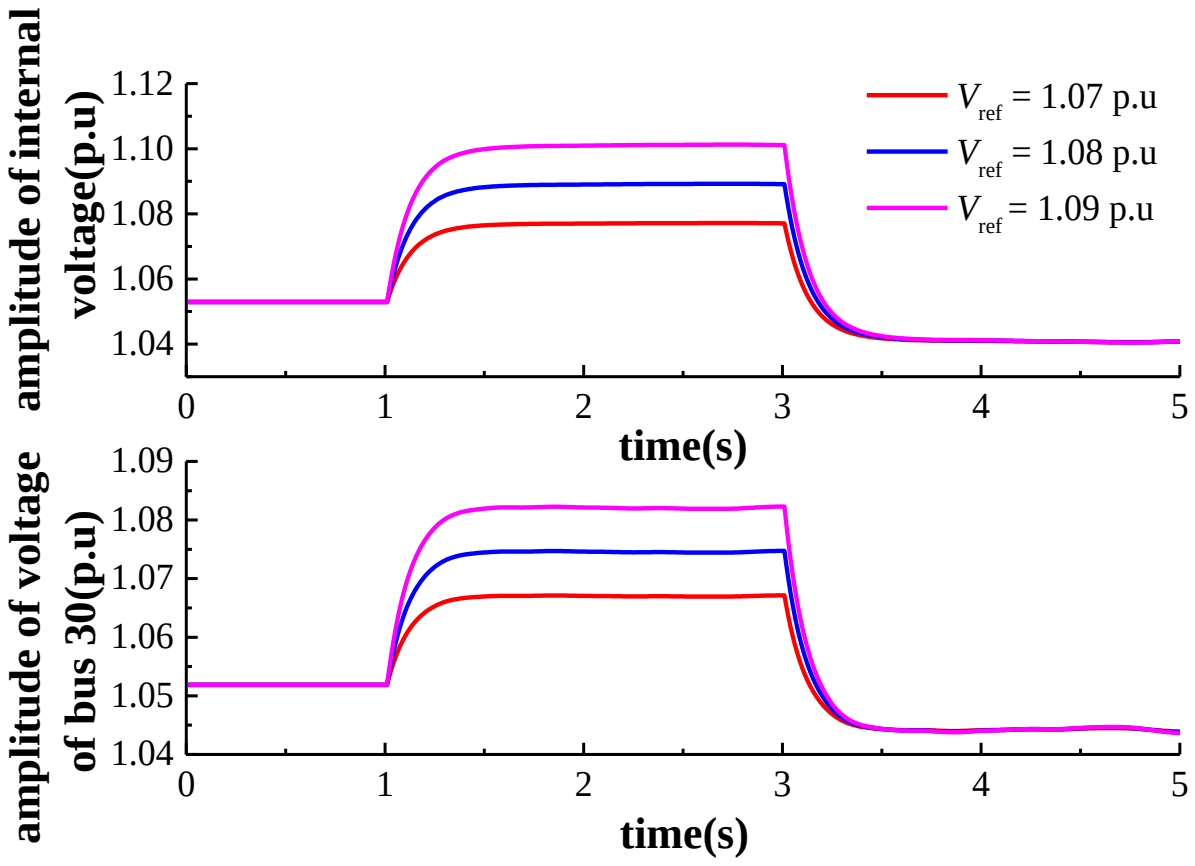

Fig. 12. The voltage trajectories of GFM converter at bus30.

It can be seen from Fig. 12 that the system voltage trajectory rises monotonically as the voltage reference value of GFM converter rises, which verifies the input-output monotonicity of the voltage trajectories. While this phenomenon is hardly surprising, the underlying principle had in fact not been fully understood. It is not difficult to see that a necessary condition under which the input-output monotonicity shown above holds true is that the system under study is stable. The exact condition ensuring the presence of the properties is not known yet, this remains as an interesting topic for future research.

*2) Voltage Trajectories under Load Shedding Controls*

It impossible to calculate the voltage derivative $\Upsilon'(\sigma)$ directly since it is a function of σ, so we calculate the minimum value instead. In Case2, the initial load at bus4 was $S = 5 + j4$ p.u and load shedding happened at $t$ = 0 s. The variations of minimum of voltage derivative $\Upsilon'(\sigma)$ for load shedding of 100 Mvar and 200 Mvar are shown in Fig. 13.

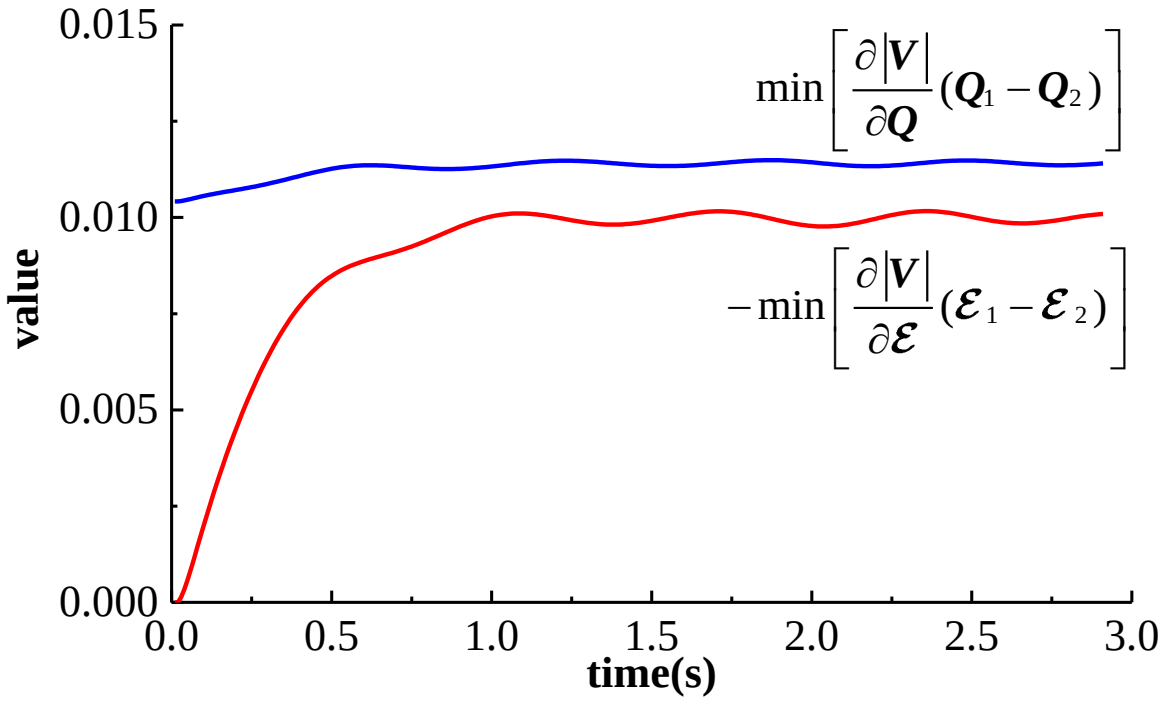

Fig. 13. The variations of $\Upsilon'(\sigma)$

It can be seen from Fig. 13 that

$$\int_0^1 \Upsilon'(\sigma)d\sigma \geq \min_{0\leq\sigma\leq1}[\Upsilon'(\sigma)](1-0) > 0 \tag{58}$$

This is consistent with the analysis in Section V, and illustrates the input-output monotonicity of load shedding control.

## VII. Conclusions

Though the input-output monotonicity properties discussed in the work have more or less been exploited by engineers in designing voltage controls, they are not fully understood. The emergence of monotone system theory provides an excellent opportunity to help look into such voltage behavior of power systems, particularly for systems with GFM converters. By leveraging the inherent mathematical characteristics of voltage subsystem model, the paper establishes the mathematical foundation of voltage control concepts from the new perspective of monotone system, and clearly elucidates conditions to guarantee control performance. The results presented also serve as a useful foundation for designing future advanced voltage controls utilizing the potentially powerful capabilities of GFM converters. Future work should focus on the application of the presented results to study the voltage dynamics of other devices such as droop-controlled GFM converters, HVDC links, and induction motors.

## Biorgraphies

**Zhenyao Li** received the B.Eng. degree from the School of Electrical and electronic Engineering, Huazhong University of Science and Technology, Wuhan, China, in 2017, and the M.Eng. degree from the College of Electrical Engineering, Zhejiang University, Hangzhou, China, in 2021. He is currently working toward the Ph.D. degree with the College of Electrical Engineering, Zhejiang University, Hangzhou, China. His research interests include stability analysis and control of grid-connected power converters.

**Shengwen Liao** received the B.S. degree in electrical engineering from North China Electric Power University, Beijing, China, in 2018. He is currently working toward the Ph.D. degree with the College of Electrical Engineering, Zhejiang University, Hangzhou, China. His research interests include voltage stability and control of AC-DC systems.

**Qian Zhang** received his B.E. and M.S. degree in Electrical Engineering from Zhejiang University, Hangzhou, China, in 2019 and 2022 respectively. He is working toward the Ph.D. degree in the Department of Electrical and Computer Engineering at Texas A&M University. His research interests include machine learning, optimization in the electricity market, and stability and control of the dynamic network.

**Xuechun Zhang** received B.Eng degree from the College of Electrical Engineering, Zhejiang University, Hangzhou, China, in 2023. She is currently working toward the Ph.D. degree with the College of Electrical Engineering, Zhejiang University, Hangzhou, China. Her research interests include the control of distributed generators and stability analysis.

**Deqiang Gan** has been with the faculty of Zhejiang University since 2002. He visited the University of Hong Kong in 2004, 2005 and 2006. Deqiang worked for ISO New England, Inc. from 1998 to 2002. He held research positions in Ibaraki University, University of Central Florida, and Cornell University from 1994 to 1998. Deqiang received a Ph.D. in Electrical Engineering from Xian Jiaotong University, China, in 1994. He served as an editor for European Transactions on Electric Power (2007-2014). His research interests are power system stability and control.